\documentclass[prl,twocolumn]{revtex4-2}
\usepackage{graphicx}
\usepackage{amsmath,amsfonts}
\usepackage{bm}

\usepackage{hyperref}

\begin{document}

\title{A gauge-invariant formulation of optical responses in superconductors}
\author{Sena Watanabe}\email{watanabe-sena397@g.ecc.u-tokyo.ac.jp}
\affiliation{Department of Applied Physics, The University of Tokyo, Tokyo 113-8656, Japan}
\author{Haruki Watanabe}\email{hwatanabe@g.ecc.u-tokyo.ac.jp}
\affiliation{Department of Applied Physics, The University of Tokyo, Tokyo 113-8656, Japan}
\date{\today}
\begin{abstract}
Superconductors are often discussed in the mean-field approximation that breaks $U(1)$ symmetry.
Since the $U(1)$ symmetry underlies the charge conservation, naive application of response theory sometimes gives results that are not gauge-invariant. 
We study the effect of vertex corrections on the electromagnetic responses of superconductors by employing the gauge-invariant formulation, called consistent fluctuations of order parameters, that manifestly satisfy the Ward identity.
For inversion broken two-band superconductors, we find that the linear and second-order optical responses are significantly affected by vertex corrections and the conductivities near the gap energy almost completely disappear.
We also investigate the linear optical responses in $d$-wave superconductors and obtained qualitatively different results  compared to the result using the solution to the Bethe-Salpeter equation. The full photon vertex used in our study, manifestly satisfies the Ward identity.
Our results demonstrate the fundamental importance of vertex corrections in exploring optical responses in superconductors.
\end{abstract}
\maketitle

\textit{Introduction---}
In superconductors, unique electromagnetic properties such as the Meissner effect and zero electrical resistance are observed~\cite{Schrieffer1999}.
Superconductivity was first observed experimentally by Kammerlingh-Onnes in 1911, but its theoretical and microscopic formulation was not made until the proposal of the Bardeen--Cooper--Schrieffer (BCS) theory~\cite{Bardeen1957}. 
In BCS theory, the $U(1)$ symmetry of the Hamiltonian is broken by the mean-field approximation.
The breaking of this symmetry underlies the unique electromagnetic properties of superconductors.

Recently, research on the optical response of various materials, including the nonlinear regime, has become increasingly active~\cite{Sipe2000,Morimoto2016, Orenstein2021}.
The optical response of quasiparticles in superconductors has also been being explored~\cite{Xu2019,Watanabe2022,Tanaka2023,Tanaka2024}.
In addition to the studies such as shift currents~\cite{Xu2019} and nonreciprocal responses~\cite{Watanabe2022}, it was found that the superconducting Berry-curvature factor is essential for the second-order optical responses~\cite{Tanaka2023}.

However, such research is based on the mean-field Hamiltonian of superconductors, and, as mentioned earlier, it lacks the $U(1)$ symmetry that underlies the conservation of the electric charge.
As a result, one often obtains results that are not apparently gauge-invariant, violating the Ward identity.
Since the gauge degree of freedom is merely a mathematical redundancy in the description, the theory must respect the gauge invariance.

This issue has long been recognized and the resolutions are known at least for the linear response of conventional superconductors.
It has been demonstrated by Nambu that the electromagnetic response kernel becomes gauge invariant if the microscopic Hamiltonian possesses $U(1)$ symmetry and many-body effects are appropriately incorporated through vertex corrections based on the Bethe-Salpeter equation~\cite{Nambu1960}.  
Using this framework, the linear electromagnetic responses of superconductors incorporating vertex corrections are investigated~\cite{Oh2024,Dai2017,Papaj2022}.
Huang and Wang examined the effect of vertex corrections of second-order responses in single-band superconductors~\cite{Huang2023}. 
In addition to Nambu's method, various other methods have been proposed~\cite{Kulik1981,Zha1995,Kosztin2000,Guo2013,Anderson2016,Boyack2016,Lutchyn2008}, and the collective mode excitations in superconductors such as Higgs mode and Leggett mode have been intensively studied~\cite{Leggett1966,Tsuji2015,Cea2016,Kamatani2022,Nagashima2024}.
However, research on collective excitations and quasiparticle excitation has been conducted separately, and there is still little knowledge about the influence of many-body effects on quasiparticle excitations, especially in the nonlinear regime.

In the case of unconventional superconductors, the research~\cite{Huang2023} has utilized the Bethe-Salpeter equation and extended it to the generalized BCS theory.
However, this treatment raises a question regarding gauge invariance, and even in linear cases, this concern remains unresolved.

In this Letter, we employ the gauge-invariant formulation of the electromagnetic response that can also be applied to unconventional superconductors.
First, we investigate the effect on quasiparticle excitations in inversion-breaking multi-band superconductors, as studied in previous research~\cite{Xu2019}.
We show that collective mode excitations exist and vertex corrections strongly suppress the linear and second-order optical conductivities originating from quasiparticle excitation.
Next, we compare the vertex corrections for anisotropic superconductors obtained using the Bethe-Salpeter equation~\cite{Huang2023} with those derived from the method proposed in this study.
We see a qualitative difference in the correction parts of the full photon vertices.

\textit{Gauge invariant response in superconductors---}
We consider spin-singlet superconductors described by the BdG Hamiltonian $\hat{H}_{\mathrm{MF}}=\sum_{\bm{k}}\hat{\psi}_{\bm{k}}^\dagger H_{\mathrm{BdG}}(\bm{k})\hat{\psi}_{\bm{k}}$,
\begin{align}
  H_{\mathrm{BdG}}(\bm{k})=\begin{pmatrix}
    H_N(\bm{k})&\Delta(\bm{k})\\
    \Delta^\dagger(\bm{k})&-H_N(\bm{-k})^T
  \end{pmatrix},
\end{align}
where $H_N(\bm{k})$ is the Hamiltonian in the normal state, $\Delta(\bm{k})$ is the gap function that satisfies $\Delta(\bm{k})^T=\Delta(-\bm{k})$, and $\hat{\psi}_{\bm{k}}$ is the Nambu spinor. 
In the velocity gauge~\cite{Parker2019}, the gauge field is introduced by the minimal coupling $H_N(\bm{k})\to H_N(\bm{k+A})$ for electrons and $H_N(-\bm{k})^T\to H_N(-\bm{k+A})^T$ for holes.
Then the bare $n$-photon vertices are obtained by derivatives of the normal Hamiltonian
\begin{align}
  \gamma^{\alpha_1\cdots\alpha_n}(k)=\prod_{i=1}^n(-\tau_3\partial_{k_i})
  \begin{pmatrix}
    H_N(\bm{k})&\\
    &-H_N(-\bm{k})^T
  \end{pmatrix},
\end{align}
where $\tau_i$ are the Pauli matrices on the Nambu basis. 
\begin{figure}[t]
  \centering
  \includegraphics[width=1\columnwidth]{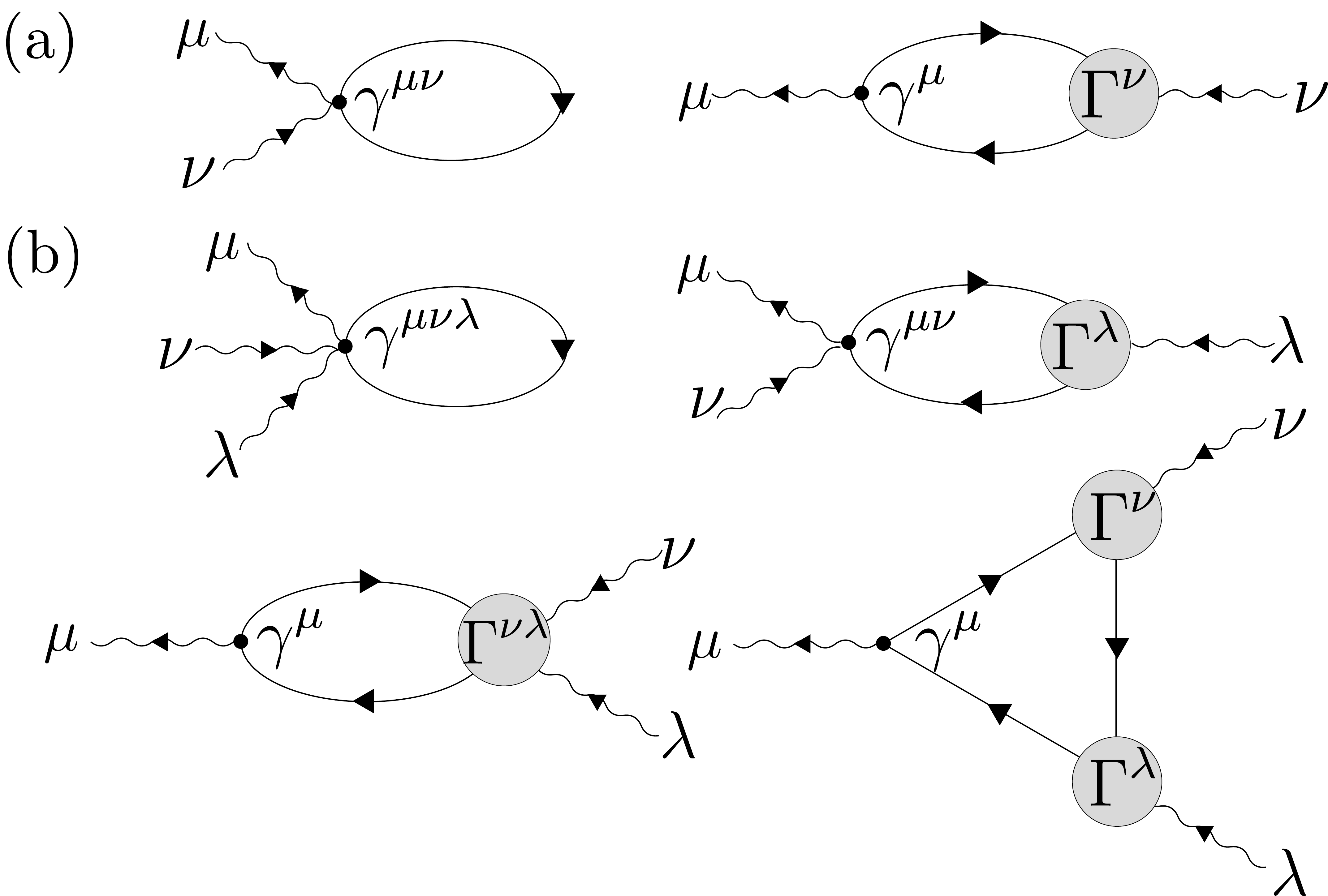}
  \caption{The diagrammatic representation of (a) the linear and (b) the second-order optical conductivities.}
  \label{fig:kernel}
\end{figure}

Since the superconductivity is a many-body phenomenon, vertex corrections have to be taken into account to obtain gauge-invariant electromagnetic responses~\cite{Nambu1960, Schrieffer1999}. 
The Green function in the presence of the gauge field is defined by $G_A(x,y)=-\langle T_\tau\hat{\psi}(x)\hat{\psi}^\dagger(y)\rangle$ where $T_\tau$ is the imaginary time-ordered product.
The full photon vertices $\Gamma^{\alpha_1\cdots\alpha_n}$ are defined by functional derivatives of the inverse of Green function with respect to the gauge field ${(-1)^{n-1}}\delta^n G_A^{-1}/\delta A_{\alpha_1}\cdots\delta A_{\alpha_n}|_{A=0}$.
The full photon vertices $\Gamma^{\alpha_1\cdots\alpha_n}$ must satisfy the Ward identity.
For example, for the one- and two-photon vertices,
\begin{align}
  \Gamma^\nu(k,q)q_\nu&=\tau_3G^{-1}(k)-G^{-1}(k+q)\tau_3,\\
  \Gamma^{\nu\lambda}(k,q,q')q'_\lambda&=\Gamma^\nu(k+q',q)\tau_3-\tau_3\Gamma^\nu(k,q),
\end{align}
where $G^{-1}(k)=k_0-H_{\mathrm{BdG}}(\bm{k})$ is the Green function in the absence of the gauge field.
Here and hereafter, summations over repeated indices are implied. 

Since the Green function satisfies the Dyson equation $G_A^{-1}(x,y)=G_{0,A}^{-1}(x,y)-\Sigma_A(x,y)$, where $G_{0,A}$ is the noninteracting Green function and $\Sigma_A$ is the self-energy, the full photon vertices $\Gamma$ can be decomposed into the sum of the bare part $\gamma$ and the correction part $\Lambda$, which is given by the functional derivative of the self-energy $\Sigma_A(x,y)$ with respect to the gauge field $A$.
The self-energy $\Sigma_A(x,y)$ is related to the superconducting gap $\Delta_{A}(\bm{r,r'})$. 
For brevity, here we assume a single band. The extension to multi-band case will be discussed later.
Since the gauge field breaks translational symmetry, one should formulate the mean-field approximation in real space.
The spin-singlet superconductivity can be realized by the microscopic electron interaction
\begin{align}
  \hat{H}_{\mathrm{int}}=-\sum_{\bm{r,r'}}\frac{g}{4}I_{\bm{r-r'}}\hat{B}_{\bm{r,r'}}^\dagger\hat{B}_{\bm{r,r'}},
\end{align}
where $\hat{B}_{\bm{r,r'}}^\dagger=\hat{c}_{\bm{r}\uparrow}^\dagger\hat{c}_{\bm{r'}\downarrow}^\dagger-\hat{c}_{\bm{r}\downarrow}^\dagger\hat{c}_{\bm{r'}\uparrow}^\dagger$ is a creation operator of a spin-singlet pair of electrons at $\bm{r}$ and $\bm{r'}$, $I_{\bm{r-r'}}>0$ is a function that specifies the spatial profile, and $g>0$ is the coupling constant~\cite{Paramekanti2000}.
If we assume $\Delta_{A}(\bm{r,r'})=-g\langle\hat{c}_{\bm{r'}\downarrow}\hat{c}_{\bm{r}\uparrow}\rangle=g\langle\hat{c}_{\bm{r'}\uparrow}\hat{c}_{\bm{r}\downarrow}\rangle$ in the mean-field approximation,
the interaction term becomes
\begin{align}
  \hat{H}_{\mathrm{int}}\simeq\sum_{\bm{r,r'}}\big(I_{\bm{r-r'}}\hat{c}_{\bm{r}\uparrow}^\dagger\hat{c}_{\bm{r'}\downarrow}^\dagger\Delta_{A}(\bm{r,r'})+h.c.\big).
\end{align}
The real part $\Delta_{1,A}$ and the imaginary part $\Delta_{2,A}$ of the gap $\Delta_{A}$ can be expressed by the Green function as
\begin{align}
  \Delta_{i,A}(x,y)=-V(x-y)\mathrm{Tr}\big[\tau_iG_A(x,y)\big]/2,\label{eq:gapeq}
\end{align}
where $V(x-y)=g\delta(x_0-y_0-0^-)I_{\bm{x-y}}$.
With these definitions, the self-energy is given by $\Sigma_A(x,y)=\sum_{i=1,2}\Delta_{i,A}(x,y)\tau_i$.

As explained above, the correction part $\Lambda_{i}^{\mu}(x,y,z)$ is given by the functional derivative $-\left.\delta\Delta_{i,A}(x,y)/\delta A_\mu(z)\right|_{A=0}$. 
Using Eq.~\eqref{eq:gapeq} and performing the Fourier transformation, we find
\begin{align}
  \Lambda_{i}^\mu(k,i\Omega)&=-\int dp\frac{V(k-p)}{2}\mathrm{Tr}\big[\tau_iG(p_0+i\Omega,\bm{p})\nonumber\\
  &\times\big(\gamma^\mu(p)+\sum_{j}\Lambda_{j}^\mu(p,i\Omega)\tau_j\big)G(p)\big].\label{eq:cfop_dwave}
\end{align}
The one-photon vertex in the ladder approximation is given by solving this integral equation.
The correction parts of the two-photon vertices $\Lambda_{i}^{\nu\lambda}(x,y,z_1,z_2)=\left.\delta^2\Delta_{i,A}(x,y)/\delta A_\nu(z_1)\delta A_\lambda(z_2)\right|_{A=0}$ can be calculated in the same manner:
\begin{align}
  &\Lambda_{i}^{\nu\lambda}(k,q_1,q_2)\nonumber\\
  &=-\int dp\frac{V(k-p)}{4}\sum_{j}\mathrm{Tr}\big[\tau_iG(p+q)\Gamma^{\nu\lambda}(p,q_1,q_2)G(p)\big]\nonumber\\
  &-\int dp\frac{V(k-p)}{2}\mathrm{Tr}\big[\tau_iG(p+q)\Gamma^\lambda(p+q,q_2)\nonumber\\
  &\qquad\times G(p+q_1)\Gamma^\nu(p,q)G(p)\big]+[(\nu,q_1)\leftrightarrow(\lambda,q_2)],\label{eq:cfop_2}
\end{align}
where $q=q_1+q_2$.
These methods to obtain the correction parts $\Lambda$ are called ``consistent fluctuations of order parameters'' (CFOP)~\cite{Kulik1981,Zha1995,Kosztin2000,Guo2013,Anderson2016,Boyack2016} and these correction parts manifestly satisfy the Ward identity~\cite{SW2024}.

After all, the optical conductivities with vertex corrections are given by
\begin{widetext}
  \begin{align}
    &\sigma^{\mu\nu}(i\Omega)=-\frac{1}{i(i\Omega)}\int dk\mathrm{Tr}\big[\gamma^{\mu\nu}(k)G(k)+\gamma^\mu(k)G(k_0+i\Omega,\bm{k})\Gamma^\nu(k,i\Omega)G(k)\big],\label{eq:lin_kernel_momentum}\\
    &\sigma^{\mu\nu\lambda}(i\Omega_1,i\Omega_2)=\frac{1}{i^2(i\Omega_1)(i\Omega_2)}\int dk\bigg(\frac{1}{2}\mathrm{Tr}\big[\gamma^{\mu\nu\lambda}(k)G(k_0)\big]+\mathrm{Tr}\big[\gamma^{\mu\nu}(k)G(k_0+i\Omega_2,\bm{k})\Gamma^\lambda(k,i\Omega_2)G(k)\big]\nonumber\\
    &\qquad+\frac{1}{2}\mathrm{Tr}\big[\gamma^\mu(k)G(k_0+i\Omega_{12},\bm{k})\Gamma^{\nu\lambda}(k,i\Omega_1,i\Omega_2)G(k)\big]\nonumber\\
    &\qquad+\mathrm{Tr}\big[\gamma^\mu(k)G(k_0+i\Omega_{12},\bm{k})\Gamma^\lambda(k,i\Omega_2)G(k_0+i\Omega_1)\Gamma^\nu(k,i\Omega_1)G(k)\big]+[(\nu,i\Omega_1)\leftrightarrow(\lambda,i\Omega_2)]\bigg),\label{eq:sec_kernel_momentum}
  \end{align}
\end{widetext}
where $\Gamma^\alpha=\gamma^\alpha+\sum_{i}\Lambda_{i}^\alpha\tau_i, \Gamma^{\alpha\beta}=\gamma^{\alpha\beta}+\sum_{i}\Lambda_{i}^{\alpha\beta}\tau_i, i\Omega_{12}=i\Omega_1+i\Omega_2$. The diagrammatic representation is presented in Fig.~\ref{fig:kernel}. 
At the end of the calculation, an analytic continuation $i\Omega_i\to \omega_i+i\eta$ with a positive infinitesimal parameter $\eta$ is performed.

\textit{Inversion-broken two-band superconductors---}
For the rest of this Letter, we apply the general formulation explained above to two important examples.
In superconductors, optical transitions between particle-hole pairs occur only when either time-reversal symmetry or inversion symmetry is broken~\cite{Ahn2021}.

\begin{figure}[t]
  \centering
  \includegraphics[width=1.0\linewidth]{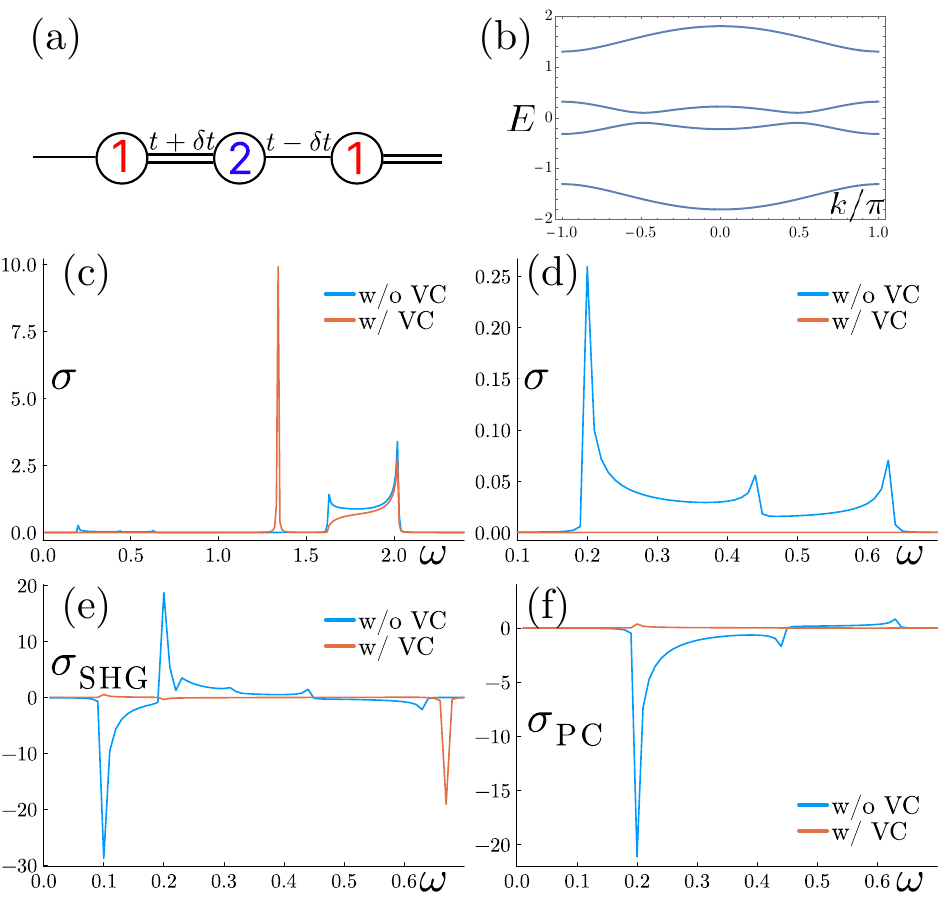}
  \caption{The linear and second-order optical responses in multi-band superconductors.
  (a) The illustration of the model in Eq.~\eqref{eq:xu_model}.
  (b) The quasi-particle spectrum. 
  (c) The real part of the linear optical conductivity $\sigma^{xx}(\omega)$ with (orange) or without (blue) vertex corrections.
  (d) Zoom-up of (c) in lower-energy regime.
  (e) The real part of the second harmonic generation $\sigma^{xxx}(\omega,\omega)$.
  (f) The real part of the photocurrent generation $\sigma^{xxx}(\omega,-\omega)$.}
  \label{fig:xu}
\end{figure}

In our first example, we break inversion symmetry by the multi-band $s$-wave gaps.
We consider a chain with sublattices $l=1,2$. As the Hamiltonian in the normal phase, we use the Rice-Mele model~\cite{Rice1982} with the zero on-site staggered potential [Fig.~\ref{fig:xu}(a)]:
\begin{align}
H_N(k)=t\cos k\sigma_1+\delta t\sin k\sigma_2-\mu,\label{eq:xu_model}
\end{align}
where $\sigma_i$ are Pauli matrices. We assume two $s$-wave superconducting gaps $\Delta(\bm{k})=\mathrm{diag}\{\Delta_1,\Delta_2\}$ as a solution to the gap equation $\Delta_l=-g_l\int dk \mathrm{Tr}\big[\tau_1E_lG(k)\big]/2$, where $[E_l]_{ab}=\delta_{la}\delta_{lb}$ are matrices acting on sublattices. 
We find $\Delta_1\neq\Delta_2$ as a result of an on-site attractive interaction $I_{\bm{r-r'},l}=\delta_{\bm{r,r'}}$ with different strength $g_l$ on each sublattice, which breaks inversion symmetry. The previous study~\cite{Xu2019} investigated the linear and second-order optical responses without vertex corrections.
Our interest is in the many-body effect on these responses.

Since the self-energy of the multi-band superconductor $\Sigma_A(x,y)=\sum_{i,l}\Delta_{i,l,A}(x,y)\tau_iE_l$ is the sum of the contributions from each sublattice,
the correction parts of vertices $\Lambda_i=\sum_l\Lambda_{il}E_l$ have the same structure.
The integral equation~\eqref{eq:cfop_dwave} can be extended to the multi-band case by the substitution $\tau_i\to \tau_iE_l$ and $V(k-p)\to V_l(k-p)$~\footnote{Details will be presented in elsewhere~\cite{SW2024}.}.
For the on-site interaction for which $V_l(k-p)=g_l$ is a constant, it is reasonable to assume that $\Lambda_{il}^\mu(k,i\Omega)$ is $k$-independent. The integral equation is reduced to the easily solvable matrix equation 
\begin{align}
  \sum_{j,l'}\bigg(\frac{2}{g_l}\delta_{il,jl'}-Q_{il,jl'}(i\Omega)\bigg)\Lambda_{jl'}^\mu(i\Omega)=Q_{il}^\mu(i\Omega),
\end{align}
where we defined correlation functions
\begin{align}
  &Q_{il}^\mu(i\Omega)=-\int dp\mathrm{Tr}\big[\tau_iE_lG(p_0+i\Omega,\bm{p})\gamma^\mu(p)G(p)\big],\\
  &Q_{il,jl'}(i\Omega)=-\int dp\mathrm{Tr}\big[\tau_iE_lG(p_0+i\Omega,\bm{p})\tau_jE_{l'}G(p)\big],
\end{align}
which are related to fluctuations of order parameters due to the applied gauge field.

For the parameters $t=1, \mu=0.8, \delta t=0.5,\Delta_1=0.15,\Delta_2=0.05,\eta=1\times 10^{-3}$, we reproduce the result of the previous study without vertex corrections.
The corresponding coupling constants are found to be $g_1\simeq 0.84,g_2\simeq 0.34$.
As shown in Fig.~\ref{fig:xu}(c), vertex corrections generally suppress the linear optical conductivity except for the new peak at $\omega\simeq 1.3$.
This is reminiscent of the corrective excitations in multi-band superconductors known as the Legget mode~\cite{Leggett1966,Kamatani2022,Nagashima2024}.

The previous study focused on low-energy quasiparticle excitations around $0.2\leq \omega\leq 0.7$ near the superconducting gap.
The calculations without vertex corrections successfully reproduce the frequency dependence reported in the previous study~\footnote{The difference in magnitude is attributed to the choice of parameter $\eta$. We also conducted the calculation with different choices of $\eta$ and the results were qualitatively unchanged. }.
With vertex corrections, however, the linear optical response in this energy regime is strongly suppressed and almost completely disappears [Fig.~\ref{fig:xu}(c)].
The vertex corrections also suppress the magnitude of second-order optical conductivity $\sigma^{xxx}(\omega_1,\omega_2)$ and may even change its sign [Fig.~\ref{fig:xu}(d,e)].

\textit{d-wave superconductors---}
Next, let us discuss single-band $d$-wave superconductors.
In square lattice, the $d$-wave pairing can be realized by the nearest-neighbor interaction  $I_{\bm{r-r'}}=\sum_{\mu=x,y}(\delta_{\bm{r,r+e_\mu}}+\delta_{\bm{r,r-e_\mu}})$.
Here, we consider the situation where an applied supercurrent induces a finite momentum $2\bm{Q}$ of Cooper pairs, which breaks inversion symmetry.
Optical responses originating from this mechanism have been studied both theoretically~\cite{Dai2017,Papaj2022,Huang2023,Crowley2022} and experimentally~\cite{Nakamura2020}.

The Hamiltonian for the normal state is
\begin{align}
  H_N(\bm{k})&=t(2-\cos(k_x+Q_x)-\cos(k_y+Q_y))-\mu.
\end{align}
The gap function is $\Delta(\bm{k})=\Delta_d\phi_d(\bm{k})$, where $\phi_d(\bm{k})=\cos k_x-\cos k_y$ is the $d$-wave form factor and $\Delta_d$ is the magnitude determined by the gap equation $\Delta_d=-g\int dp\sum_p\mathrm{Tr}\big[\tau_1G(p)\big]\phi_d(\bm{p})/2$. 

For this model, the previous study obtained the one-photon vertex by solving the Bethe-Salpeter equation~\cite{Huang2023}
\begin{align}
  \Gamma^\nu(k,i\Omega)&=\gamma^\nu(k)+g\phi_d(\bm{k})\int dp\phi_d(\bm{p})\tau_3\nonumber\\
  &\times G(p_0+i\Omega,\bm{p})\Gamma^\nu(p,i\Omega)G(p)\tau_3\label{eq:bse_gBCS}.
\end{align}
In this study, the correction part is expanded as $\Lambda^\mu(k,q)=\sum_{i=0}^3\Lambda_i^\mu(k,q)\tau_i$ including the diagonal components $\tau_0,\tau_3$.
Also, $\Lambda_i^\mu(k,i\Omega)$ is assumed to respect the $d$-wave symmetry, $\Lambda_i^\mu(k,i\Omega)=\phi_d(\bm{k})\Lambda_i^\mu(i\Omega)$.
However, we found that the vertex obtained this way does not necessarily satisfy the Ward identity.
This is because the proof of the Ward identity does not go through when the interaction is not of $V(k-p)$ form.

\begin{figure}[t]
  \centering
  \includegraphics[width=1\linewidth]{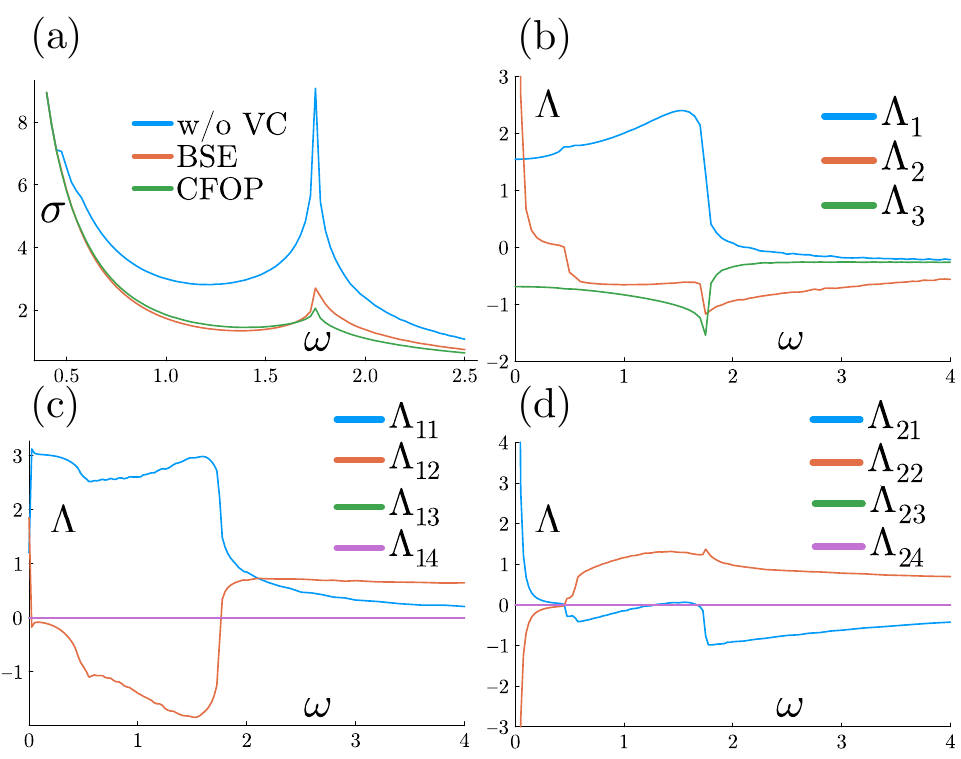}
  \caption{(a) The real part of optical conductivity $\sigma^{xx}(\omega)$ with (orange, green) or without (blue) vertex corrections. The orange corresponds to the result of Eq.~\eqref{eq:bse_gBCS}, and the green corresponds to that of Eq.~\eqref{eq:cfop_dwave}.
  (b) The real part of the solution $\Lambda_i^x(\omega)$ to Eq.~\eqref{eq:bse_gBCS}.
  (c) The real part of the $\Lambda_{1n}^x(\omega)$ and (d) the real part of $\Lambda_{2n}^x(\omega)$.}
  \label{fig:dwave_VC}
\end{figure}

We thus solve Eq.~\eqref{eq:cfop_dwave}, instead of Eq.~\eqref{eq:bse_gBCS}, so that the Ward identity is manifest. 
In our method, the diagonal components $\tau_0,\tau_3$ are absent.
To this end, it is beneficial to separate $k$-dependence of $V(k-p)$ as $V(k-p)=2g(\cos(k_x-p_x)+\cos(k_y-p_y))=2g\sum_{n=1}^4f_n(\bm{k})f_n(\bm{p})$ where $f_1(\bm{k})=\cos k_x$, $f_2(\bm{k})=\cos k_y$, $f_3(\bm{k})=\sin k_x$, and $f_4(\bm{k})=\sin k_y.$
Similarly, we expand the solution to the integral equation as $\Lambda_i^\mu(k,i\Omega)=\sum_{n=1}^4f_n(\bm{k})\Lambda_{in}^\mu(i\Omega)$, where $\Lambda_{in}^\mu(i\Omega)$ satisfies
\begin{align}
  \Lambda_{in}^\mu(i\Omega)&=-g\int dpf_n(\bm{p})\mathrm{Tr}\big[\tau_iG(p_0+i\Omega,\bm{p})&\nonumber\\
  &\times\big(\gamma^\mu(p)+\sum_{j,m}\tau_jf_m(p)\big)G(p)\big]\Lambda_{jm}^\mu(i\Omega).\label{eq:cfop_mat}
\end{align}
This integral equation can be rewritten into the form of the matrix equation
\begin{align}
  \sum_{j,m}\bigg(\frac{1}{g}\delta_{in,jm}-Q_{in,jm}(i\Omega)\bigg)\Lambda_{jm}^\mu(i\Omega)=Q_{in}^\mu(i\Omega),
\end{align}
where
\begin{align}
&Q_{in}^\mu(i\Omega)=-\int dpf_n(p)\mathrm{Tr}\big[\tau_iG(p_0+i\Omega,\bm{p})\gamma^\mu(p)G(p)\big],\\
&Q_{in,jm}(i\Omega)=-\int dpf_n(p)f_m(p)\mathrm{Tr}\big[\tau_iG(p_0+i\Omega,\bm{p})\tau_jG(p)\big].
\end{align}

Following the previous study~\cite{Huang2023}, we set the parameters as $t=1.0\times 10^2, \mu=9.0\times 10^1,\Delta_d=2.3\times 10^1,\eta=3\times 10^{-2}$ and assume the Cooper pair momentum along $x$-direction $\bm{Q}=Q_x\bm{e_x}$ with $Q_x=0.07$.  
The corresponding coupling constant is $g\simeq 1.8\times 10^2$.

We compare the linear optical response in two methods in Fig.~\ref{fig:dwave_VC} (a), which are at least quantitatively different although the difference is not significant.
The vertex corrections based on the Bethe-Salpeter equation reproduce the results reported in the previous study [Fig.~\ref{fig:dwave_VC}(b)], for which the $d$-wave symmetry is assumed.
In contrast, the correction part obtained in this study does not possess $d$-wave symmetry since $\Lambda_{i1}^\mu(i\Omega)\neq-\Lambda_{i2}^\mu(i\Omega)$ as shown in Fig.~\ref{fig:dwave_VC}(c,d).  Our result reflects the  breaking of $C_4$ rotational symmetry by the momentum of the Cooper pairs.

\textit{Conclusion---}
In this Letter, we employed the gauge-invariant formulation of electromagnetic responses and numerically investigated the impact of vertex corrections on optical responses in superconductors.
In inversion-broken multi-band superconductors, the peak corresponding to the quasiparticle excitation near gap energy is strongly suppressed and almost completely disappears. In the case of $d$-wave superconductors, the photon vertex used in this research has the advantage of explicitly satisfying the Ward identity.
Our result implies the importance of treating optical responses in superconductors in a gauge-invariant manner.

\begin{acknowledgments}
We thank Y.~Yanase, A.~Daido, H.~Tanaka, and T.~Morimoto for useful discussions.
We were informed that H.~Tanaka and Y.~Yanase were working on a similar problem but in a different situation and they independently obtained formulas that correspond to Eqs.~\eqref{eq:cfop_dwave} and ~\eqref{eq:cfop_2}.
The work of S.W. is supported by World-leading Innovative Graduate Study Program for Materials Research, Information, and Technology~(MERIT-WINGS) of the University of Tokyo.
The work of H.W. is supported by JSPS KAKENHI Grant No.~JP24K00541.

\end{acknowledgments}

\bibliography{bcs_library}

\end{document}